\documentclass[%
 aip,
 amsmath,amssymb,
 reprint,%
]{revtex4-1}

\usepackage{graphicx}
\usepackage{dcolumn}
\usepackage{bm}

\usepackage[utf8]{inputenc}
\usepackage[T1]{fontenc}
\usepackage{mathptmx}
\usepackage{etoolbox}

\makeatletter
\def\@email#1#2{%
 \endgroup
 \patchcmd{\titleblock@produce}
  {\frontmatter@RRAPformat}
  {\frontmatter@RRAPformat{\produce@RRAP{*#1\href{mailto:#2}{#2}}}\frontmatter@RRAPformat}
  {}{}
}%
\makeatother
\begin{document}


\title[Radio Frequency Electrical Resistance Measurement under Destructive Pulsed Magnetic Fields]{Radio Frequency Electrical Resistance Measurement under Destructive Pulsed Magnetic Fields}

\author{T. Shitaokoshi}
\affiliation{
Institute for Solid State Physics, The University of Tokyo, Kashiwa, Chiba 277-8581, Japan
}
 
\author{S. Kawachi}
\affiliation{ 
Graduate School of Science, University of Hyogo, Ako, Hyogo 678-1297, Japan
}

\author{T. Nomura}
\affiliation{
Institute for Solid State Physics, The University of Tokyo, Kashiwa, Chiba 277-8581, Japan
}
\affiliation{
 School of Engineering, Tokyo Denki University, Adachi, Tokyo 120-8551, Japan
}

\author{F. F. Balakirev}
\affiliation{
National High Magnetic Field Laboratory, Los Alamos National Laboratory, Los Alamos, NM 87545, United States
}

\author{Y. Kohama}
\email{ykohama@issp.u-tokyo.ac.jp}
\affiliation{
Institute for Solid State Physics, The University of Tokyo, Kashiwa, Chiba 277-8581, Japan
}

\date{\today}

\begin{abstract}
We developed a resistance measurement using radio frequency reflection to investigate the electrical transport characteristics under destructive pulsed magnetic fields above 100 T.  A homemade flexible printed circuit for a sample stage reduced the noise caused by the induced voltage from the pulsed magnetic fields, improving the accuracy of the measurements of the reflected waves. From the obtained reflectance data, the absolute value of the magnetoresistance was successfully determined by using a phase analysis with admittance charts. These developments enable more accurate and comprehensive measurements of electrical resistance in pulsed magnetic fields.
\end{abstract}

\maketitle

\section{Introduction}

In the study of metals and their transport properties, the measurement of electrical resistivity in high magnetic fields plays a crucial role. Conventional approaches for measuring magnetoresistance have relied on standard alternating current (AC) techniques, typically conducted at magnetic fields below 100 T, generated by non-destructive pulsed magnets. However, performing magnetoresistance measurements in higher magnetic fields exceeding 100 T has been a significant challenge, requiring the use of destructive pulse magnets such as a single-turn coil (STC). The STC, while capable of generating high magnetic fields up to \(\sim\)200 T, \cite{Herlach_1973} has an extremely short field duration of several microseconds accompanied by significant electrical noise due to its destructive nature, which is unsuitable for conventional resistance measurements.

In previous attempts to measure the electrical resistivity of metallic samples in strong magnetic fields generated by the STC, several approaches were taken, including the four-terminal, \cite{Takamasu_1996,Nakagawa_1998,Miura_2002} millimeter-wave transmission, \cite{Shimamoto_1998} radio frequency (RF) transmission, \cite{Sakakibara_1989,Sekitani_2003,Sekitani_2007} and self-resonant RF reflection measurements. \cite{Nakamura_2018} However, most of these works faced challenges due to the rapid sweep rate of the STC ( \(\sim 10^8\) T/s), which induces a huge voltage (\(\sim 10^2\) V) in the circuit in proportion to its cross-sectional area perpendicular to the magnetic field. This significant voltage prevents accurate measurements in these experiments. 

In recent years, an alternative approach has been developed, \cite{Kohama_2020} which employs a flexible printed circuit (FPC) that is laid parallel to the magnetic field and has an extremely small cross-sectional area perpendicular to the field. This configuration minimizes the induced voltage to an optimal level, allowing it to be used as a voltage source for resistivity measurements. However, in this technique, the induced voltage becomes zero near the peak of the pulsed magnetic field, which means it is impossible to measure resistance in the maximum field region.

Among the previous approaches, the RF impedance measurement, \cite{Saito_2003,Inokuchi_2004,Imamura_2006,Hanzawa_2022} which utilizes the relationship between the reflectance of RF waves and electrical impedance, offers the advantage of being able to perform measurements at the peak of the magnetic field. In the conventional RF impedance measurements, a directional coupler was used to separate RF waves. \cite{Saito_2003,Inokuchi_2004,Imamura_2006} However, the weak coupling between the input and the coupled port of the directional coupler resulted in insufficient separation of the reflected waves. This led to weak signals and a low signal-to-noise (S/N) ratio, compromising the accuracy and reliability of the measurements. In addition, the transmission line used in the conventional setup consisted of a twisted pair of copper wires, which was not fully impedance-matched and prevented RF propagation. In addition, the small loops formed in the twisted pair resulted in excess induced voltage, introducing noise into the measurements. Furthermore, the appropriate analysis of the RF impedance was not applied, resulting in inaccurate resistance values. Especially, the reflection coefficient, originally a complex number, was treated as a real number, neglecting the phase information. Therefore, there was a need for enhanced analytical techniques to fully utilize the complex nature of the reflection coefficient and improve the accuracy of the resistance measurements.

  In this paper, we will introduce a setup for the RF impedance measurement utilizing a circulator instead of a directional coupler for the RF wave separation and an FPC for noise reduction. Furthermore, we present a novel analytical method for obtaining magnetoresistance data, where the reflection coefficient is treated as a complex number, incorporating the use of an admittance chart. In the following sections, we will explain the measurement setup, theoretical foundations, and practical data analysis using the result obtained with a non-destructive pulse magnet. We will also present the results obtained with an STC. Comparative studies with conventional methods will be included to highlight the advantages and limitations of our novel approach.

\begin{figure*}
    \centering
    \includegraphics[width=0.9\linewidth]{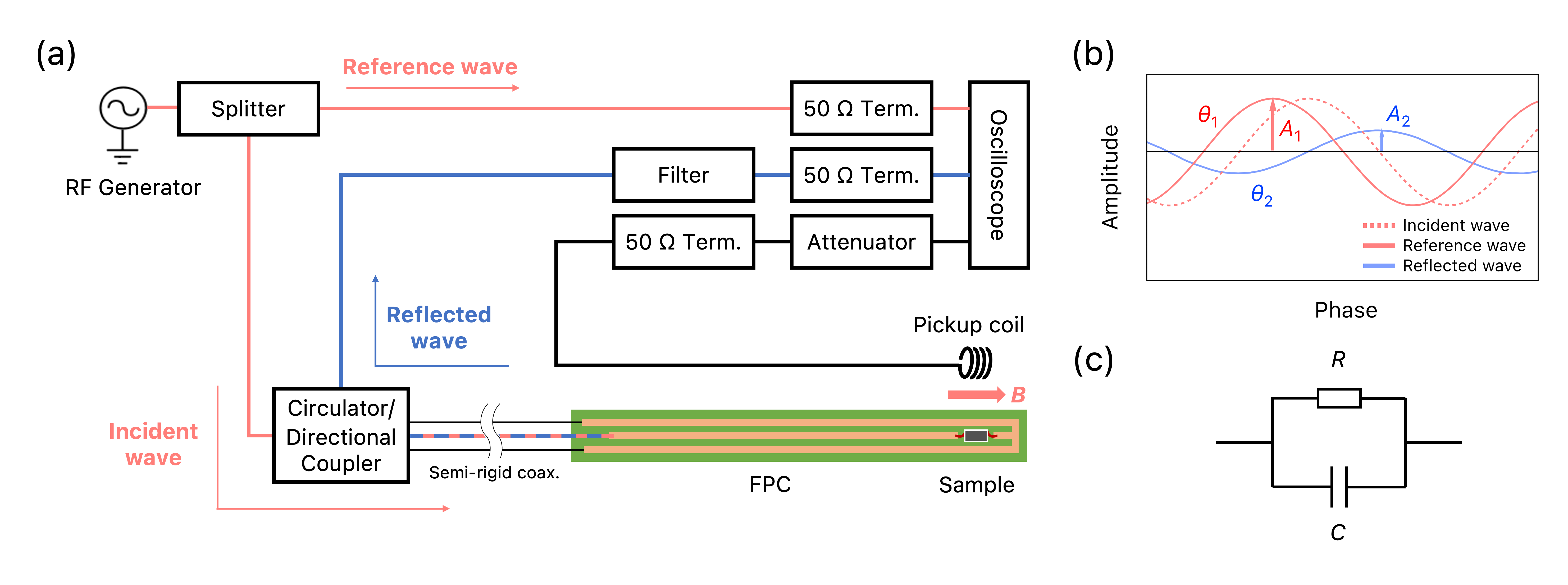}
    \caption{
(a) Block diagram of the RF measurements. Signal from an RF generator is split in two and one half of the signal is captured as a reference wave, and the other half is sent via a circulator or directional coupler to a conductive sample mounted on an FPC. The reflected signal from the sample is separated at the circulator or directional coupler and captured after passing through the band-pass filter. 50 \(\Omega\) terminators are attached to each input. 
(b) Schematic diagram of the amplitudes and phases of the incident, reflected and reference wave. The amplitude and the phase of the reflected wave change due to the reflection at the sample. The phase of the reference wave is shifted from that of the incident wave according to the cable length.
(c) An equivalent circuit around the sample. \(R\) and \(C\) represent the resistance of the sample and the parasitic capacitance associated with the wires and other components, respectively.
}
    \label{fig:circuit}
\end{figure*}

\section{Method}

\subsection{Experiment}

Figure \ref{fig:circuit}(a) shows the block diagram of an RF impedance-measurement system which consists of an RF generator, oscilloscope, splitter, circulator, terminator, filter, attenuator, pickup coil, and FPC. RF sine wave of 125 or 250 MHz was generated by the combination of a bandpass filter with Crystek CPRO33 or CRBSCS-01 clock oscillator, respectively, and signals between 130-140 MHz were generated by Stanford Research Systems SG382 signal generator. The signals were split in two by Mini-Circuits ZFSC-2-1-S+ splitter. Half of the signals were measured as a reference wave by Teledyne LeCroy HDO4104A oscilloscope, while the other half were transmitted to the conducting sample on the FPC via the circulator. Admotech ADC012CSLH(50)-B circulator was used for the measurements between 125--140 MHz, and ADC028CSLH(80) for the measurements at 250 MHz. These results were compared with those at 125 MHz, where the circulator was replaced by a directional coupler (Mini-Circuits ZFDC-10-1+).

The reflected waves from the sample were separated by the circulator or the directional coupler and recorded by the oscilloscope after passing through a bandpass filter. To prevent unwanted RF reflections at the input of the oscilloscope and to protect the oscilloscope from the spike noise induced by the field generation, external 50 \(\Omega\) terminators were connected prior to the oscilloscope, and input impedances of the oscilloscope were set to 1 M\(\Omega\). The reflected waves and the reference waves were numerically treated with the lock-in technique, and the reflectance \(\Gamma=A_2/A_1\) and the phase shift \(\theta=\theta_2-\theta_1\) were obtained, where \(A_1\) and \(A_2\) are the amplitudes of the reference and reflected waves and \(\theta_1\) and \(\theta_2\) are the phases of the reference and reflected waves, respectively (See Fig. \ref{fig:circuit}(b)). 

Measurements at low temperatures and under pulsed magnetic fields were conducted in a plastic or metal cryostat using a measurement probe consisting of a 10-cm-long FPC, a 90-cm-long semi-rigid coaxial cable, and a fiber-reinforced plastic rod. The FPC fabricated by Taiyo Industrial consisted of a three-line microstrip array \cite{Pavlidis_1976,Tripathi_1977} connected to the semi-rigid coaxial cable with a solder at one of the ends. The three lines were connected at the opposite end, and the central line had a gap to accommodate a metallic sample (Fig. \ref{fig:circuit}(a)). The two outer lines were grounded, functioning as electromagnetic shields. Each line had a width of 100 \(\mathrm{\mu}\)m and they were spaced 50 \(\mathrm{\mu}\)m apart from each other. The FPC and semi-rigid coaxial cable were glued on the fiber-reinforced plastic rod. The characteristic impedance of the FPC (\(Z_0\)) is designed to be around 50 \(\Omega\). The magnetic fields were generated by a non-destructive pulse magnet and an STC. The time durations of these field pulses were 38 ms and 7 \(\mathrm{\mu}\)s, respectively. The field strength was calculated by integrating the voltage induced in the pickup coil. 

One of the test samples used for this work was kish graphite, which was micro-processed using focused ion beam (FIB) technology. The sample dimensions after processing were approximately 100 \(\mathrm{\mu}\)m x 25 \(\mathrm{\mu}\)m x 10 \(\mathrm{\mu}\)m. The graphite sample was mounted on the FPC using gold wires and silver epoxy. Measurements in the non-destructive pulse magnetic fields were conducted at 4.2 K using liquid helium. The other test sample was \(\mathrm{YBa_2Cu_3O_7}\) (YBCO) thin film grown on MgO substrate which was purchased from CERACO Ltd. The YBCO film has a slight excess of copper and shows the superconducting transition at \(\sim\)86 K. The YBCO film had a thickness of 500 nm and was cut into the dimensions of 1500 \(\mathrm{\mu}\)m x 100 \(\mathrm{\mu}\)m. Two different film samples were prepared from the same batch of YBCO film and used for the experiments at the STC (sample 1) and non-destructive magnet (sample 2). The measurements were conducted between 77.4--87.5 K using liquid nitrogen. In addition to the above test samples, small chip resistors ranging from 30 to 100 \(\Omega\) were also mounted on the FPC, and the performance of the probe was checked under zero magnetic field at 77.4 K.

To check the validity of our technique, standard 2-terminal resistance measurements were also performed using the same probe at low frequency (50 kHz) with a non-destructive pulsed magnet. The circulator or the directional coupler was removed and a shunt resistor was connected in series with the sample. The voltage between each end of the sample (\(V\)) and the shunt resistor (\(V_{\mathrm{shunt}}\)) was measured. The 2-terminal resistance of the sample was calculated by \(R = R_{\mathrm{shunt}} V/V_{\mathrm{shunt}}\), where \(R\) and \(R_{\mathrm{shunt}}\) are the resistance of the sample and shunt resistor, respectively.

\subsection{Analysis}

The actual reflectance \(\Gamma_\mathrm{S}\) at the sample, which is the ratio between the amplitudes of the incident and reflected wave, and the phase shift \(\theta\) can be transformed into polar coordinates to define the complex reflection coefficient 
\begin{equation}\label{eq:polar}
    \rho=\Gamma_\mathrm{S} e^{i\theta} = u + iv,
\end{equation}
where \(u\) and \(v\) are the real and imaginary parts of the complex coefficient, respectively.
In this analysis, the sample connected to the FPC with wires was treated as a parallel circuit consisting of a real resistance \(R\) and a parasitic capacitance \(C\)  \cite{Xie_1996} (See Fig. \ref{fig:circuit}(c)), and the admittance of this parallel \(RC\) circuit was considered.
The relationship between the normalized admittance \(y = YZ_0\) (\(Y\): admittance and \(Z_0\): characteristic impedance) and the complex reflection coefficient \(\rho\) is given by, \cite{Basu_2023}
\begin{equation}\label{eq:adm}
y = \frac{1-\rho}{1+\rho}.    
\end{equation}
Complex numbers \(y = g + ib\) and  \(\rho = u + iv\) are substituted into this equation, where \(g\) and \(b\) are the normalized conductance and susceptance, respectively. By comparing the real and imaginary parts of Eq. \eqref{eq:adm}, one can obtain
\begin{equation}\label{eq:conduct}
    g = \frac{1-u^2-v^2}{(1+u)^2+v^2}    
\end{equation}
\begin{equation}\label{eq:suscept}
    b = -\frac{2v}{(1+u)^2+v^2}.    
\end{equation}
The relationship obtained from Eqs. \eqref{eq:conduct} and \eqref{eq:suscept} represent circles in the \(uv\)-plane, the so-called admittance chart. The admittance chart is a graphical calculator used for electronic circuit design in the field of RF engineering, forming a complementary pair with the Smith chart. \cite{Basu_2023,Mizuhashi_1937,Smith_1939,Smith_1941} When \(\rho\) is plotted on the \(uv\)-plane along with an admittance chart, it tells the conductance and susceptance of the sample. This means that, by measuring the variation of \(\rho\) with the magnetic field and plotting it with an admittance chart, one can convert it to the field dependence of the conductance and susceptance.

\section{Results}

\subsection{Measurements with non-destructive magnet}

\begin{figure*}
    \centering
    \includegraphics[width=0.65\linewidth]{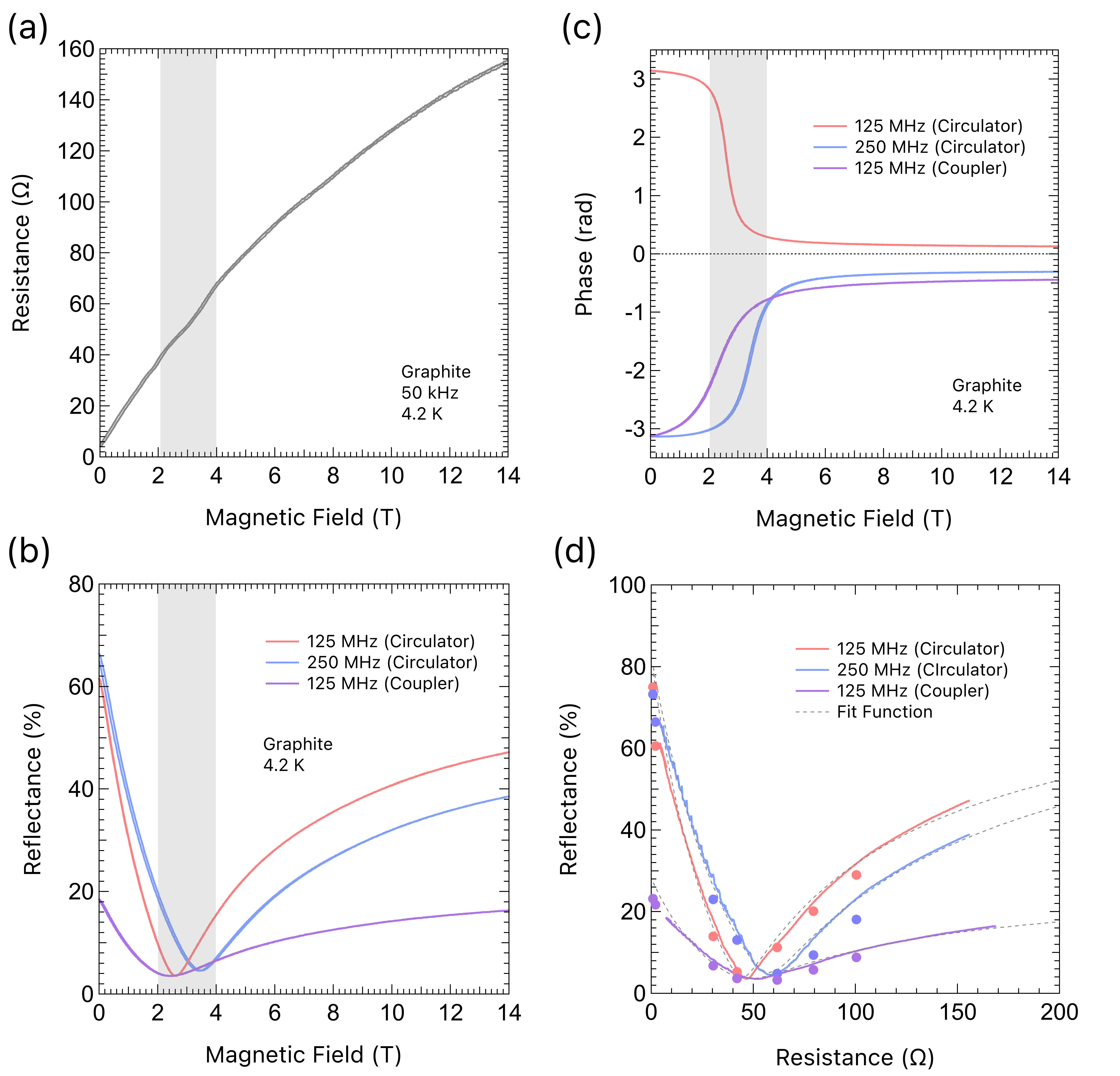}
    \caption{(a) Two-terminal resistance of FIB-processed graphite measured by AC technique in the non-destructive pulsed magnetic field. (b) Reflectance and (c) phase shift obtained by RF measurements. Reflectance shows the ratio of the amplitude of the reflected wave to that of the reference wave, and phase shift represents the difference of the phase of those waves. A constant background caused by the phase rotation during propagation in the coaxial cable has been subtracted from the measured phase. The gray hatched area shows the magnetic field region of 2--4 T where the magnetoresistance of graphite reaches around 50 \(\Omega\), which is the characteristic impedance of the transmission line. (d) Reflectance vs resistance plot. The red solid line shows the data at 125 MHz obtained with the circulator. The blue solid line shows the data at 250 MHz with the circulator. The purple solid line shows the data obtained at 125 MHz with the directional coupler. The filled circles represent the data obtained from the chip resistors. The Gray dashed line shows the fit curve obtained by Eq. \eqref{eq:fit2}.}
    \label{fig:RFdata}
\end{figure*}

Figure \ref{fig:RFdata}(a) shows the magnetoresistance of FIB-processed graphite measured by the standard 2-terminal technique at a frequency of 50 kHz. The magnetic field was generated by the non-destructive pulsed magnet and the sample was immersed into liquid helium at 4.2 K. Graphite exhibits a large magnetoresistance in relatively weak magnetic fields, which is appropriate for testing the validity of the analysis proposed in the previous section. 

Figure \ref{fig:RFdata}(b) displays the RF reflectance \(\Gamma\) measured at 125 and 250 MHz, using the circulator or directional coupler. In all setups, a dip was observed in \(\Gamma\), where the magnetoresistance of the sample reaches the characteristic impedance \(Z_0\) (\(\sim\) 50 \(\Omega\)) and the reflection is minimized due to the impedance matching (gray hatched area). The magnetic field where the dip appears depends on the frequency or the measurement setup because of the variation of \(Z_0\). The use of the circulator instead of the directional coupler results in lower losses and larger reflection signals. 

Figure \ref{fig:RFdata}(c) illustrates the field dependence of the phase shift \(\theta(B)\) obtained from RF measurements at 125 and 250 MHz. The phase shift is caused by the reflection from the sample and depend on its admittance. Around zero magnetic field, where the sample resistance is nearly zero and the system can be regarded as a short circuit, \(\theta(B)\) is close to \(\pm\pi\), which is equivalent to the fixed-end reflection. In the higher field, where the sample resistance reaches several hundreds of ohm, which can be considered as the free-end reflection, \(\theta(B)\) approaches zero, At around fields where the sample resistance is close to \(Z_0\) (impedance matching), \(\theta(B)\) changes drastically. The qualitative difference in the sign of \(\theta(B)\) is attributed to the variation in the susceptance of the circuit. Note that the wavelength of RF waves used here is around 1--2 m, and the total length of the transmission line is \(\sim\)10 m, which results in a phase offset of 20--10 \(\pi\). To eliminate this background, a constant value is subtracted from the raw phase data as discussed below.

In Fig. \ref{fig:RFdata}(d), the solid curves represent the relationship between the reflectance and the resistance of graphite derived from Figs. \ref{fig:RFdata}(a) and \ref{fig:RFdata}(b). The filled circles show the data obtained from chip resistors.
The positions of the dips in the curves provide information on \(Z_0\)  of the measurement system at each setup. The dashed lines show the fit function of the obtained \(\Gamma\) vs \(R\) curves. The fitting procedure is performed as follows. As in the Eq. \eqref{eq:polar},  \(\Gamma_\mathrm{S}\) is the absolute value of \(\rho\), which can be transformed as,
\begin{equation}\label{eq:fit1}
    \Gamma_\mathrm{S}=|\rho|=\left|\frac{1-y}{1+y}\right|
    =\left|\frac{1-g-ib}{1+g+ib}\right|
    =\left|\frac{R-Z_0-iRb}{R+Z_0+iRb}\right| .
\end{equation}
Therefore, the observed \(\Gamma\) vs \(R\) curves can be fitted by using the following function
\begin{equation}\label{eq:fit2}
    \Gamma = A\Gamma_\mathrm{S}= A\sqrt{\frac{(R-Z_0)^2+(Rb)^2}{(R+Z_0)^2+(Rb)^2}},
\end{equation}
where \(A\) is a parameter that relates to several factors such as the attenuation in the transmission line and coupling constant between the ports of the splitter/circulator/directional coupler.
Table \ref{tab:fit} presents the fit parameters obtained from the above fits.

\begin{table}
\caption{\label{tab:fit} Fitting parameters of \(\Gamma\) vs \(R\) data obtained by Eq. \eqref{eq:fit2}. \(A\) is the maximum reflectance, \(Z_0\) is the characteristic impedance, and \(|b|\) is the absolute value of susceptance.}
\begin{ruledtabular}
\begin{tabular}{lccc}
Setup&\(A\)&\(Z_0\)&\(|b|\)\\
\hline
125 MHz (Circulator)                  &0.79   &43.8 \(\Omega\)    &0.11\\
250 MHz (Circulator)                  &0.83   &58.2 \(\Omega\)    &0.12\\
125 MHz (Directional Coupler) &0.28   &49.9 \(\Omega\)    &0.26\\
\end{tabular}
\end{ruledtabular}
\end{table}

\begin{figure}
    \centering
    \includegraphics[width=1\linewidth]{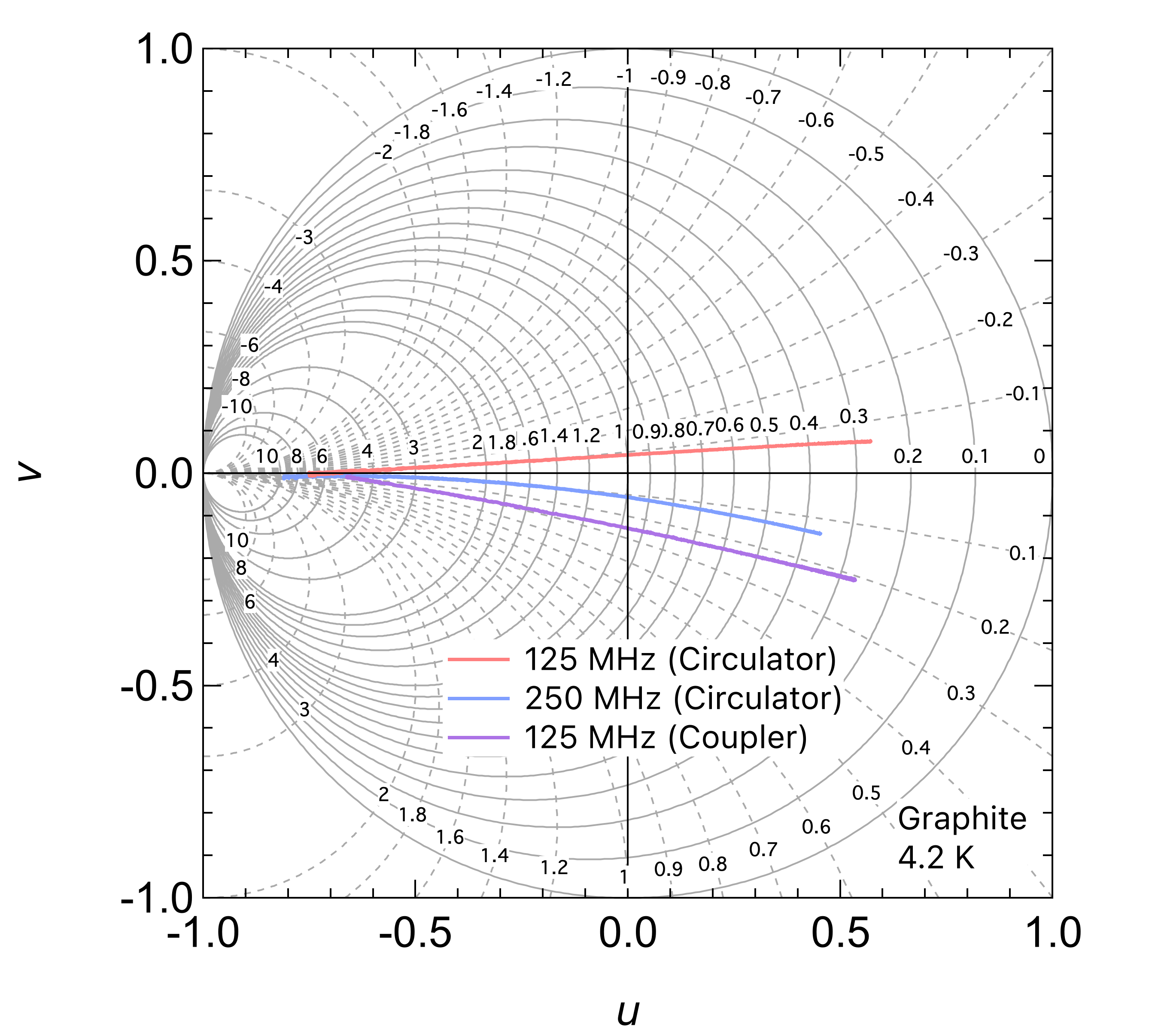}
    \caption{\(\Gamma/A\)  and \(\theta\) are plotted on \(uv\)-plane using polar coordinates along with an admittance chart. The red line shows the data at 125 MHz using circulator, while the blue line represents the data at 250 MHz. The purple line corresponds to the data at 125 MHz using directional coupler. Solid and dashed gray lines show the constant conductance circle and the constant susceptance circle of the admittance chart, respectively. The numbers on the circles represent the value of conductance and susceptance of each line, which are normalized by the fitted values of \(Z_0\).}
    \label{fig:admchart}
\end{figure}

In Fig. \ref{fig:admchart}, the reflectance at the sample \(\Gamma_\mathrm{S} = \Gamma/A\) and phase \(\theta\) are plotted on the polar coordinates, which is identical to plotting the complex reflection coefficient \(\rho\) on the \(uv\)-plane. An admittance chart is overlaid on the plot, which provides a graphical representation of \(\rho\) in terms of conductance and susceptance. The observed value of the phase shift contains the influence of the extrinsic phase rotations within the transmission line and electric parts of the present setup. Since the degree of the extrinsic phase rotations depends on the length of the cable and measurement frequency, a constant background is subtracted from the raw data to obtain \(\theta(B)\) in Fig. \ref{fig:RFdata}(c), which is equivalent to rotating the plot around the origin in Fig. \ref{fig:admchart}. In the present case, the conductance of the graphite sample changes as a function of the field, while the capacitance of the graphite sample should keep a constant value. Therefore, the constant background of \(\theta\) is chosen to align the trace of \(\rho\) on the constant susceptance circle of the admittance chart, enabling to follow the change of conductance. Regardless of the measurement frequencies, \(\rho\) changes from the left side to the right side of the admittance chart with increasing magnetic fields, which shows a decrease in conductance or equivalently an increase in resistance of the graphite sample. The values of the constant susceptance circles, along which \(\rho\) is aligned, depend on the frequency and the measurement setup. This might be attributed to the differences in the parasitic inductance or capacitance of the measurement circuit.

\begin{figure*}
    \centering
    \includegraphics[width=0.7\linewidth]{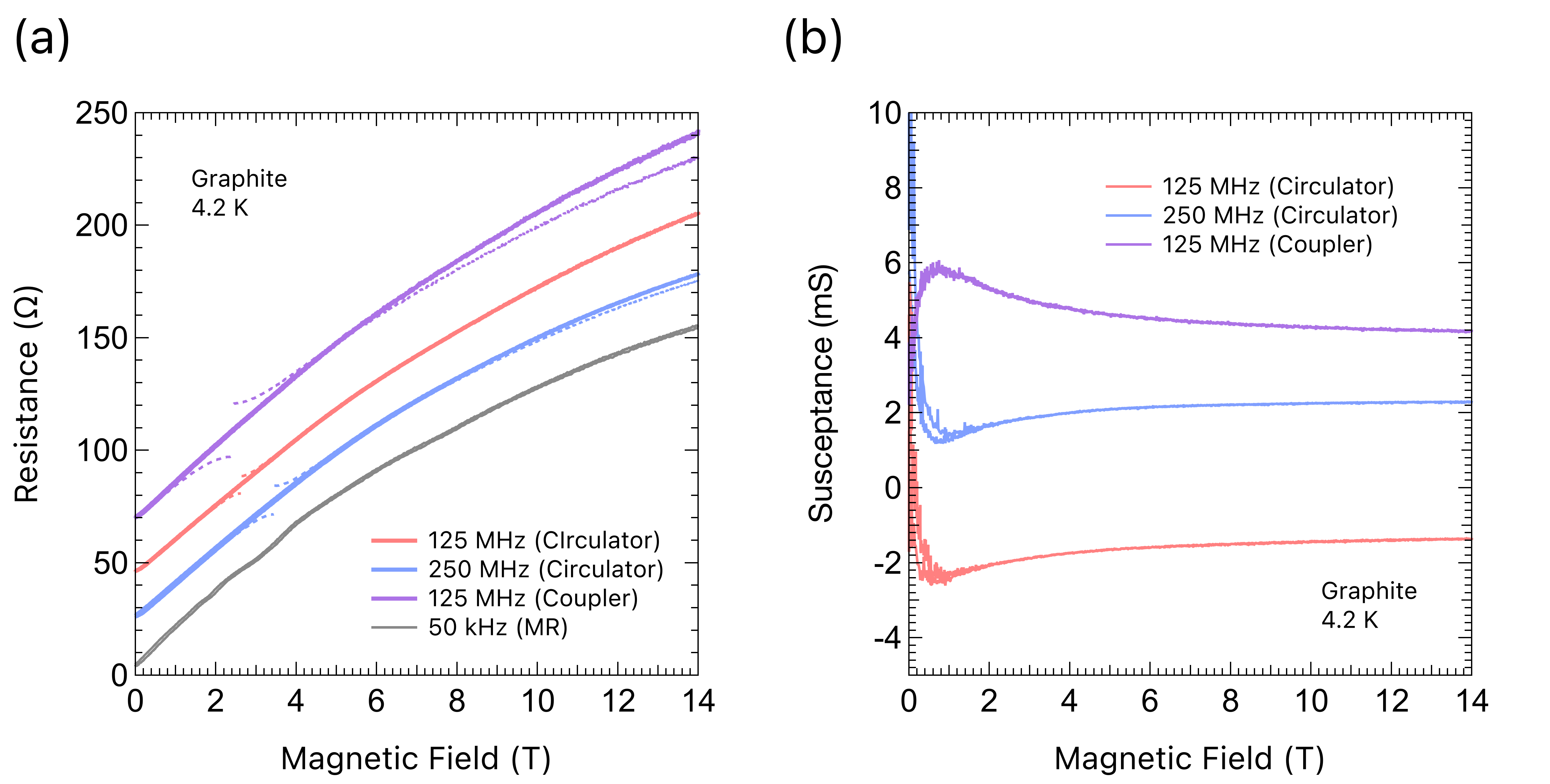}
    \caption{(a) Magnetoresistance of FIB-processed graphite obtained from RF impedance and 2-terminal AC resistance measurements. The purple solid line shows the data at 125 MHz with the directional coupler. The red solid line shows the data at 125 MHz with the circulator. The blue solid line shows the data at 250 MHz with the circulator. The black line represents the AC magnetoresistance at 50 kHz. Dashed lines represent magnetoresistance calculated using the conventional method. Each line is offset by 20 \(\Omega\). (b) Magnetic field dependence of susceptance obtained from RF impedance measurements.}
    \label{fig:resist_suscept}
\end{figure*}

Figure \ref{fig:resist_suscept} (a, b) shows the field dependence of resistance and susceptance of graphite, respectively, which are calculated from the \(\rho\) curves plotted in Fig. \ref{fig:admchart}. By applying Eqs. \eqref{eq:conduct} and \eqref{eq:suscept}, they are converted into normalized conductance and susceptance. The resistance is obtained by \(R = Z_0/g\), and the susceptance \(B\) is calculated by \(B = b/Z_0\). The calculated resistance curves exhibit good agreements with the 2-terminal magnetoresistance measured at the low frequency of 50 kHz. By ignoring the imaginary part of \(\rho\) and substituting it in Eq. \eqref{eq:adm} as in the previous reports, \cite{Saito_2003,Inokuchi_2004,Imamura_2006} the resistance curves were also evaluated as seen in the dashed curves in Fig. \ref{fig:resist_suscept}(a). When comparing the magnetoresistance curves obtained by Eqs. \eqref{eq:conduct} and \eqref{eq:suscept} (solid curves) with those obtained by Eq. \eqref{eq:adm}, it is clear that the magnetoresistance calculated by Eq. \eqref{eq:adm} exhibits discontinuities when the magnetoresistance crosses the region of characteristic impedance. We found that the field dependence of susceptance remains nearly constant above 2 T, while the magnetoresistance shows an extremely large increase from 5 \(\Omega\) (0 T) to 150 \(\Omega\) (14 T). The fluctuations of 6 mS observed near zero magnetic field are due to the poor sensitivity on the left side of the admittance chart, where the constant susceptance circles are densely gathered.

\subsection{Measurements with STC}

RF impedance measurements at 140 MHz were performed in STC, and standard 2-terminal resistance measurements at 50 kHz were conducted in the non-destructive magnet. The YBCO sample was immersed in liquid nitrogen. In the non-destructive magnet, magnetic fields up to 57 T were generated, and the magnetoresistance was measured in helium gas cooled by liquid nitrogen.

\begin{figure}
    \centering
    \includegraphics[width=1\linewidth]{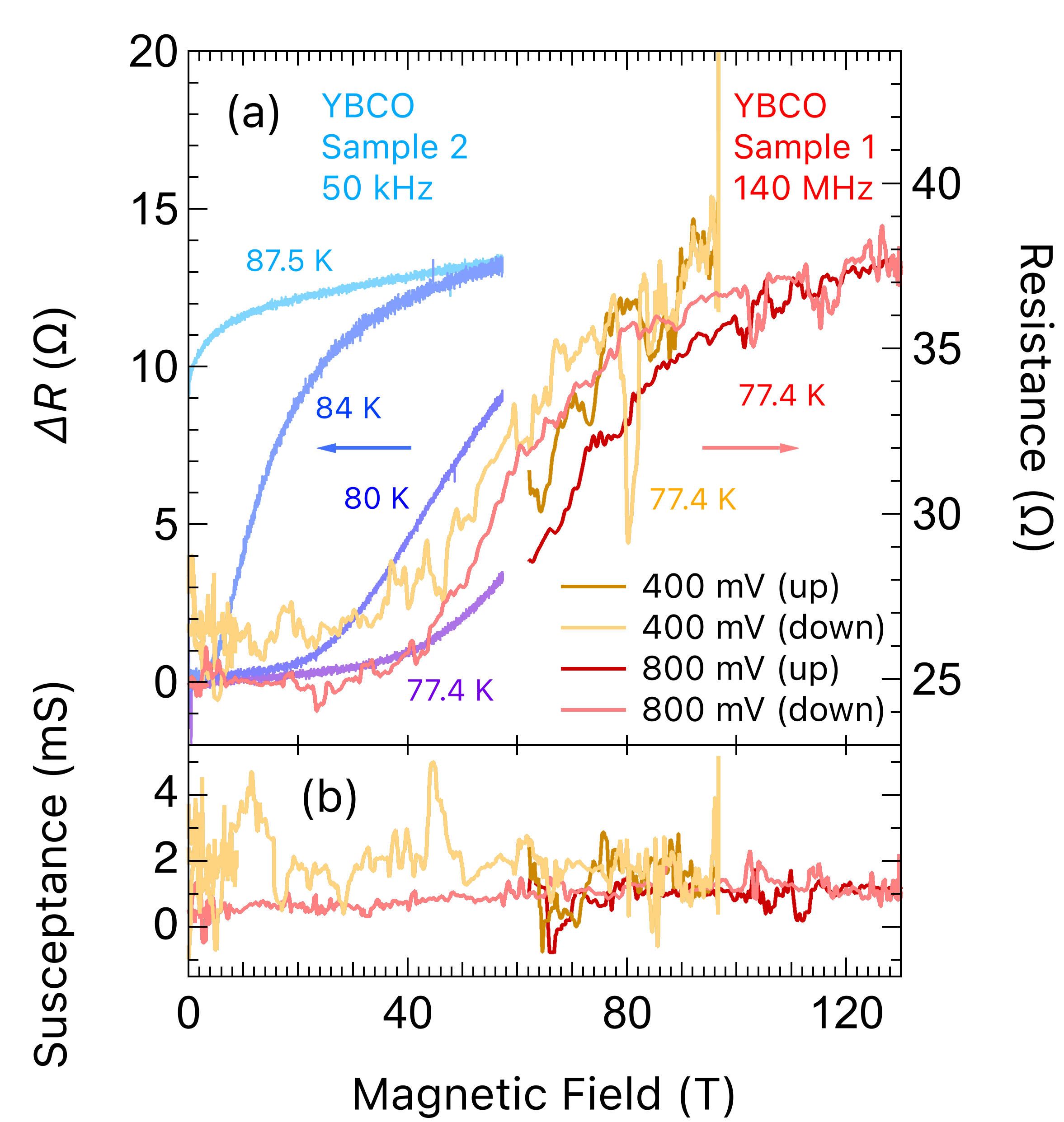}
    \caption{(a) Magnetic field dependence of resistance of YBCO superconducting film. Yellow and red lines show the calculated resistance of sample 1 obtained from 140 MHz RF impedance measurements at 77.4 K with STC (right axis). The dark and light yellow lines represent the data obtained with the RF source of 400 mV in the increasing and decreasing magnetic field, respectively. The dark and light red lines depict the data taken with the RF source of 800 mV in the increasing and decreasing field, respectively. The contribution of the contact resistance was not subtracted. Cyan, blue, cobalt, and purple lines show the change of magnetoresistance \(\Delta R\) of sample 2, obtained from 50 kHz standard 2-terminal measurements with the non-destructive magnet at 87.5, 84, 80, and 77.4 K, respectively (left axis). 
   The contribution of the contact resistance was subtracted for the estimation of \(\Delta R\). (b) Field dependence of susceptance obtained from RF measurements of sample 1 with STC.}
    \label{fig:YBCO_STC}
\end{figure}

In Fig. \ref{fig:YBCO_STC}(a), the converted magnetoresistance obtained through RF impedance measurements in STC at 77.4 K is presented (sample 1, right axis), along with the magnetoresistance change of YBCO obtained in the non-destructive magnet at several temperatures (sample 2, left axis). For the STC data, the signal within approximately 1 \(\mathrm{\mu s}\) after the pulse generation was excluded from the analysis due to the noise accompanied by the field generation. The RF data was analyzed using an admittance chart and converted into resistance and susceptance, where the attenuation coefficient \(A\) and characteristic impedance \(Z_0\) were set to be 0.8 and 44 \(\Omega\), respectively. Two STC experiments were performed with maximum magnetic fields of 100 and 130 T, and RF waves with a frequency of 140 MHz and peak-to-peak amplitudes of 400 and 800 mV were used, respectively. The data obtained with the RF amplitude of 800 mV shows a better S/N ratio. The slight difference in the zero-field resistance in each experiment is probably due to the change in contact resistance. In both experiments, the middle points of the superconducting to the metal transition are located at \(\sim\) 65 T. These magnetoresistance data obtained by the RF impedance technique agree with the data taken in the non-destructive magnet at 77.4 K, although there is a slight hysteresis in the data taken with STC. One possibility is that the hysteresis is due to the intrinsic character of the YBCO superconductor. \cite{Marcon_1991} Another possibility is the eddy current heating, which occurs in metallic parts of the probe or a metallic sample itself, namely wires, electrodes, and the YBCO sample. Even if the eddy current heating is the main reason for the hysteresis, the increased temperature during the experiment is less than 80 K, considering the magnetoresistance data taken with the non-destructive magnet.

Figure \ref{fig:YBCO_STC}(b) illustrates the field dependence of susceptance obtained at STC. The susceptance remains nearly constant throughout the measurements. Since the YBCO film is a metallic material, no change in the capacitance component is expected. This confirms that the present analysis correctly treats the background of the phase.

\section{Summary}

We applied the RF impedance technique to perform magnetoresistance measurements in ultrahigh magnetic fields above 100 T. The use of a circulator instead of a directional coupler enhanced the separation of reflected waves. Replacing twisted pairs of copper wires with FPC and semi-rigid coaxial cables reduced the voltage induced by the pulsed field and improved the S/N ratio. Additionally, by treating the reflection coefficient as a complex number and utilizing an admittance chart, we achieved the separation of the magnetoresistance and the magnetosusceptance. These advancements provide more precise and comprehensive insights into the electrical properties of materials under magnetic fields.

\begin{acknowledgments}
The authors are grateful to S. Peng and Y. H. Matsuda. This work was partly supported by a JSPS KAKENHI Grant Numbers 22H00104 and the New Energy and Industrial Technology Development Organization (NEDO) JPNP20004. The National High Magnetic Field Laboratory is supported by the National Science Foundation through NSF/DMR-2128556* and the State of Florida.
\end{acknowledgments}

\section*{Author Declarations}

\subsection*{Conflict of Interest}

The authors have no conflicts to disclose.

\section*{Data Availability Statement}

The data that support the findings of this study are available from the corresponding author upon reasonable request.

\section*{References}

\nocite{*}
\bibliography{reference}

\end{document}